\documentclass[runningheads]{llncs}
\usepackage{cite}
\usepackage{amsmath,amssymb,amsfonts}
\usepackage{graphicx}
\usepackage{xcolor}

\begin{document}

\title{Characterizing behavioral trends in a community driven discussion platform}

\author{Sachin Thukral\inst{1} \and  Arnab Chatterjee\inst{1}  
\and Hardik Meisheri\inst{1}  
\and Tushar Kataria\inst{2} \and Aman Agarwal\inst{2} 
\and Ishan Verma\inst{1}  \and Lipika Dey\inst{1}}

\authorrunning{S. Thukral et al.}

\institute{TCS Research 
\email{sachi.2,arnab.chatterjee4,hardik.meisheri,ishan.verma,lipika.dey@tcs.com}
\and IIIT Delhi, New Delhi, India\\
\email{tushar15184,aman15012@iiitd.ac.in}}

\maketitle
\begin{abstract}

This article presents a  systematic analysis of the patterns of behavior of individuals as well as groups observed in community-driven platforms for discussion like Reddit, 
where users usually exchange information and viewpoints on their topics of interest. We perform a statistical analysis of the behavior of posts and model the users' interactions around them. 
A platform like Reddit which has grown exponentially, starting from a very small community to one of the largest social networks, with its large user base and popularity harboring a variety of 
behavior of users in terms of their activity. Our work provides interesting insights about a huge number of inactive posts which fail to attract attention despite their authors exhibiting \textit{Cyborg-like} behavior to attract attention. We also observe short-lived yet extremely active posts emulate a phenomenon like \textit{Mayfly Buzz}. A method is presented, to study the activity around posts which are highly active, to determine the presence of \textit{Limelight} hogging activity. 
We also present a systematic analysis to study the presence of controversies in posts.
We analyzed data from two periods of one-year duration but separated by few years in time, to understand how social media has evolved through the years. 

\end{abstract}


\section{Introduction}

The  availability of massive amounts of data from electronic footprints of social behavior of humans for a variety of online social networks has triggered a lot of research and its applications. 
Tools from various disciplines have come together and is currently popular as computational social science~\cite{lazer2009computational}. The contemporary approach to social network analysis has improved upon the standard, classical approaches~\cite{wasserman1994social} of sociologists, and the present interests span across various disciplines like market intelligence, operations research, survey science, as well as statistical computing.
The studies of social network data not only reveals the structure of the connections between its individual components, including their strong and weak ties, and their dynamics, but also the possible reasons as to why those structure and dynamics are prevalent.

A typical social network can be conceived as a multidimensional graph where different elements like  users, posts, comments, etc. are the nodes and the links define their interactions. Usual measurable quantities like the lifespan of a post, the average number of posts per unit time etc. reveal aggregate behavior of users across the entire social media platform. User comments across posts render the interactivity flavor, where user behavior can be segregated according to volume of comments or time span of interaction. The layer of the number of distinct users who are involved, opens a scope to differentiate the behavior of users in terms of impact of the post and its reachability. 

Access to huge amounts of data facilitates a rigorous statistical analysis, which can be combined with behavioral studies that can bring out interesting spatial and temporal features, which provides interesting insights. In this article, we study evolution patterns of the posts over time, based on user interactions with the posts and group them into further different categories. We also categorized posts according to user interaction patterns that emerge around them. We present methods to determine the focal points of interactions. Further, we present methods to identify the behavioral trends exhibited by the users to make their posts popular. Additionally, we also discuss methods to analyze the presence of controversial posts and comments, even before getting into the text content.

In \textit{Reddit.com}, users share content in the form of text posts, links and images, which can be voted up/down by other users, where from further discussions may emerge. Posts span over a variety of topics -- news, movies, science, music, books, video games, fitness, food, image-sharing, etc. They are organized by subject under user-created \textit{subreddits}, which provide further opportunities for fostering discussion, raising attention and publicity for causes.
Reddit is known for its open nature and harbors diverse user community across demographics and subcultures, who generate its content, there is also moderation of posts due to various reasons.

In this article, we gathered insights about where, when and by whom the content is being created in the community as a whole. The study of evolution patterns helps us to understand the characteristics of posts which get large number of responses. Studying these characteristics from an author's perspective gives us an indication on which authors are relatively more reliable in spreading information. Similarly, identifying  the focal points of a long discussion may lead to understanding of popular opinions. These markers of behavioral trends can be used as cues in applications like placement of advertisement, summarizing of viral/popular topics from varied perspectives, the half life of information spread, etc. 

Social media is being increasingly used for sharing important information across individuals and collaboration, even within enterprises. Understanding human behavior within them and being able to characterize them, as well as to understand the dynamics of interaction within group of users turns out to be an challenging task. 
For example, in the organization to which most of the authors of this study are affiliated to, more than $400,000$ employees engage in at least two organization specific, closed social networks serving different purposes. Analysis of the temporal patterns and the group dynamics presented in our work are important aspects which can not only aid in understanding the different categories of users, but also identify the information needs and push the right content or advertisement for the right group at the right time. The similarity of patterns observed over multiple data sources prove that user behaviors are fairly similar across social platforms in the same domain.

This article is essentially an extension of our recent paper~\cite{thukral2018analyzing} where we presented the primary analysis of behavioral trends observed in Reddit.
The rest of the article is organized as follows: We present the earlier related work in Section~\ref{sec:related}. A brief description of the data considered for our study is given in Section~\ref{sec:data}. 
Analysis of evolution patterns of the posts is reported in Section~\ref{sec:evolution_patterns}. Section~\ref{sec:interaction} reports the interaction dynamics, while Section~\ref{sec:author} discusses the behavior of authors over the space. Section~\ref{sec:controversy} discusses methods to identify the presence of controversial content in posts and comments.
Finally, the entire analysis is summarized along with the inferences in Section~\ref{sec:conclusion}.

\section{Related Work}
\label{sec:related}
There have been several studies on social media dynamics from various perspectives. Researchers have examined the structure of the comment threads by analyzing the radial tree representation of thread hierarchies~\cite{gomez2008statistical}. 
Researchers have also studied the behavioral aspects of users by crowd-sourcing information from experiments performed on the platform. One such study reports how individuals consume information through social news websites, contributing to their ranking systems. A study of the browsing and rating pattern reported that most users hardly read the article that they vote on, and in fact $73\%$ of posts were rated before even viewing the content~\cite{glenski2017consumers}.  While user interactions (likes, votes, clicks, and views) serve as a proxy for the content's quality, popularity, or news-worthiness, predicting user behavior was found to be relatively easy~\cite{glenski2017predicting}. The voting pattern in the Reddit~\cite{mills2011researching} has been studied to analyze the upvoting of posts from a new page to the front page and behavior of users towards some posts which are getting positive or negative votes.They have studied the posts mentioning Wikileaks and Fox News and to see the impact of negative voting on them, although working on only one month of data. One study related to rating effect on posts and comments~\cite{glenski2017rating} revealed that random rating manipulations on posts and comments led to significant changes in downstream ratings leading to significantly different final outcomes -- positive herding effects for positive treatments on posts, increasing final ratings on the average, but not for positive treatments on comments, while negative herding effects for negative treatments on posts and comments, decreasing the final ratings on average. Another exploratory study~\cite{weninger2013exploration} on the dynamics of discussion threads found topical hierarchy in discussion threads, and how it is  possible to use them to enhance Web search. A study on `social roles' of users~\cite{buntain2014identifying} found that the typical ``answer person'' role is quite prominent, while such individual users are not active beyond one particular subreddit.
In another study, authors have used the volume of comments a blog post receives as a proxy of popularity to model the relationship with the text~\cite{Worthy_of_comment}. Authors have used several regression models to predict the volume of comments given in the text. This analysis is quite restricted in terms of the scale of the dataset, limiting to political posts and only three websites which amount to four thousand posts. While the content analysis is most intuitive, it does not provide richer analysis. Text content shared over social media is usually noisy, full of non-standard grammar and spelling, often cryptic and uninformative to the outsider from the community. When one adds the scale of today's social media data, it is computationally non-viable to have content analysis over the whole corpus.

Most of the studies reported till date have performed analysis on a subset of data by restricting themselves to a limited number of posts, comments, top users, subreddits, etc., while we use two separate sets each of which are complete data for one year period. To the best of our knowledge, only very few researchers have used complete data for analysis. In Ref.~\cite{singer2014evolution}, authors have presented evolution analysis over five years of subreddits with respect to text, images, and links though they have only considered posts and not analyzed comments. Ref.~\cite{missing_data} has reported the effect of missing data and its implications over the Reddit corpus taken from 2005 to 2016.

{  The study of controversies in social media platforms is important in several contexts. The earliest papers that dealt with the issue mainly focused on controversies arising in political conversations, in a typical setting of U.S. Presidential elections, from political news and blogs as well as posts in Twitter.
There have been reports of high correlation between negative affect and biased language with controversial issues~\cite{mejova2014controversy}. In a study of Twitter data, there are reports of highly segregated partisan structure from the retweet graph, with sparse connectivity between the opposite polarities~\cite{conover2011political}. There exists a body of literature like the above, which make use of the interaction graph structure to uncover and eventually quantify controversies (See also Ref.~\cite{garimella2018quantifying}). However, a text based, sentiment analysis adds to the overall picture, and is also attempted in a few studies~\cite{mejova2014controversy,choi2010identifying}.
We take a rather simple, rigorous statistical approach, to lay the foundation for future text-based analysis of content. The Reddit corpus is usually devoid of controversial text, as they are removed by moderators, and hence the remaining corpus around the deleted content requires detailed analysis, if one aims at finding indicators of forthcoming controversies.
}

\section{Data description}
\label{sec:data}

\subsection{Terminologies}
The following are the  terminologies that will be frequently used throughout the article:
\begin{itemize}
	\item A Reddit \textbf{Post} is text, link or an image posted by a registered user. 
	
	\item \textbf{Comment} is a response to an active post on Reddit. It is a direct response to the post or a response to any comment made on a post, thus creating a nested structure of a tree graph with possibly any number of offsprings at any level.
	
	\item \textbf{Author} is a registered user on the platform who has at least one post or comment. 
	
	\item \textbf{Score} is the difference between the number of upvotes and downvotes.

\end{itemize}

\subsection{Definitions}
We define the following quantities, which we will use in the rest of the article:
\begin{itemize}
	\item \textbf{Age} is the time difference between the last comment on the post and creation of the post, measured in seconds (unless otherwise mentioned).
	\item \textbf{Effective Comments} are the total number of comments on a post made by users other than the author of the post.
    \item \textbf{Automoderator} Reddit's Official bot.
    \item { \textbf{Deleted Author} Authors whose unique identity is absent while posts or comments are present in the metadata. }	
\end{itemize}

\subsection{Data}
We use two separate data sets of Reddit~\cite{Reddit_dump}, in order to see if the data exhibits any qualitative changes along with the quantitative changes,over a gap of few years -- 
\begin{itemize}
\item 
Period I: 1 January 2008 - 31 December 2008,
\item 
 Period II: 1 August 2014 - 31 July 2015.
\end{itemize}
We considered posts and comments for entire periods of one year each along with associated meta-data like the title of a post, time of post/comment, subreddit topic, parent post/comment id, etc. We have considered only those comments which were made on a post within the periods of study. We also neglected the comments made during Period I to posts created before Period I. The same was followed while analyzing content for Period II as well.  Table~\ref{tab:basic_data} summarizes the basic statistics of the data. It is important to note that after September 2015, there was a change in the number of fields that were being provided by Reddit API. So, in order to maintain consistency, we have used the data till July 2015.
We analyzed the data using parallel computation on a Hadoop setup.
\begin{table}[h!]
	\begin{center}
		\caption{2008 Data Table}
		\label{tab:basic_data}
        \resizebox{\columnwidth}{!}{
		\begin{tabular}{|l|r|r|}
        \hline
            & Period I & Period II \\
			\hline
			Number of Posts & 2,523,761 & 63,118,764 \\
			\hline
			Posts with deleted authors & 425,770 (16.87$\%$) & 12,346,042 (19.56$\%$)\\
			\hline
			Posts with zero comments & 1,536,962 & 23,417,869\\
			\hline
			Posts with one comment & 591,489 & 9,011,332\\
			\hline
			Number of Comments & 7,242,871 & 613,385,507\\
			\hline
			Number of Comments on posts of the period & 7,224,539 & 608,654,680\\
			\hline
            Number of Disconnected Posts & 219 (0.009$\%$) & 1,380 (0.002$\%$)\\
            \hline
            Number of Removed Comments & 355 (0.004 $\%$) & 248,493 (0.04$\%$)\\
            \hline
		\end{tabular}
  }
	\end{center}
\end{table}

\begin{table}[h!]
	\begin{center}
		\caption{Used Data Variables for posts and comments}
		\label{tab:data_Variables}
		\begin{tabular}{|l|l|}
			\hline
			\textbf{Posts} & \textbf{Comments}\\
			\hline
			author & author\\
            created utc & created utc\\
             & link id\\
            name & name\\
            number of comments & parent id\\
			\hline
		\end{tabular}
	\end{center}
\end{table}
Table~\ref{tab:data_Variables} shows set of variables from the available metadata for posts and comments that are used for our study. 
We have not used score in our analyses, except for the case of cyborgs.

\begin{figure}[t]
\centering	\includegraphics[width=0.48\linewidth]{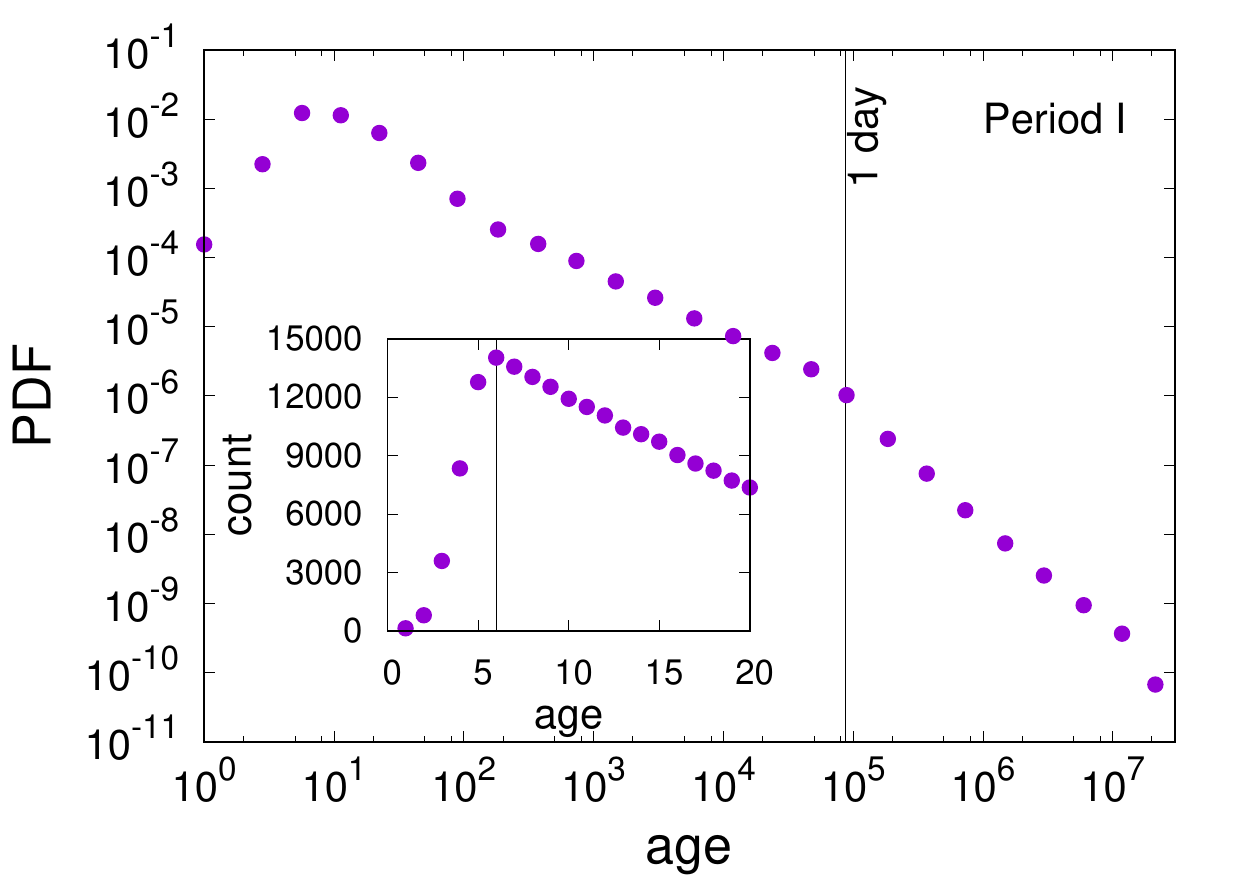}
 \centering   \includegraphics[width=0.48\linewidth]{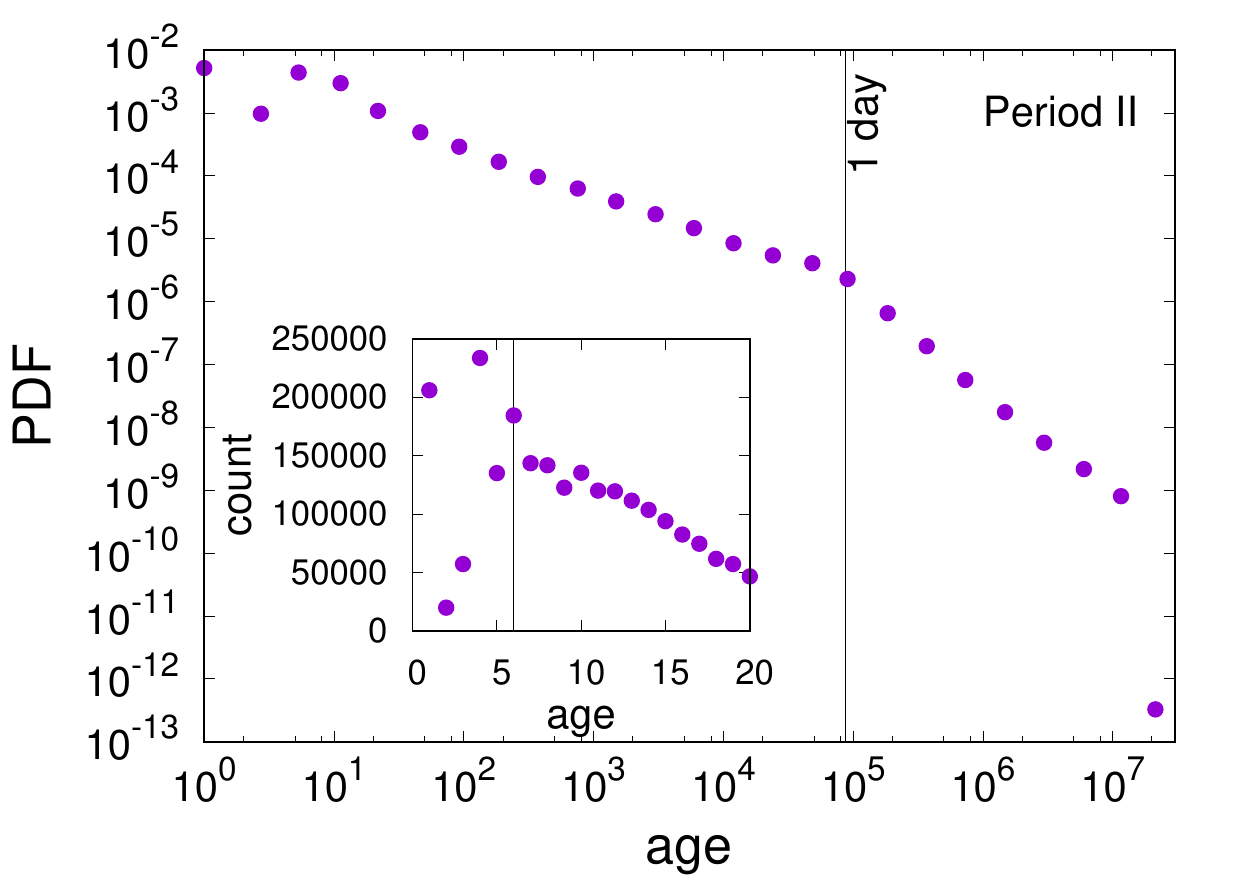}
	\caption{The PDF of the age of a post (seconds) over the entire range of age for Periods I and II in double logarithmic scale. The nature of the probability distribution shows a prominent change of slope around $\approx 1$ day, indicating that a large fraction of posts become inactive after that time.
		The insets show the corresponding histograms of the age distribution at small values of age, in double linear scale. Prominent peak around $6$ seconds for Period I and at values less than that for Period II are notable. 
	}
	\label{fig:age_posts}
\end{figure}

\section{Analysis of post evolution patterns}
\label{sec:evolution_patterns}
To analyze the evolution of the posts, we calculate the age and number of comments for each post.

\subsection{Mayfly Buzz}
The probability density function (PDF) of the ages of all posts (Figure~\ref{fig:age_posts}) has a most likely value at $6$ seconds for Period I, whereas the equivalent peak is smeared across values less than $6$ seconds for Period II.
Also, we observe that there is a shift in slope around the age of 1 day, after which, the PDF declines more quickly, suggesting that more posts tend to become inactive after a day. In fact, $ 88.6\%$ of posts die within a day in period I and $ 71.1\%$ of posts in period II. We call this post behavior \textit{Mayfly Buzz }, which resonates with the idea of creating a day buzz. As seen in other social networking platforms, activity usually dies after a very short period of time. It is interesting to see a similar behavior on Reddit, which is a discussion platform as opposed to a microblogging site such as Twitter~\cite{kwak2010twitter}, where a post's average age is longer than a tweet.  

\subsection{Cyborg-like behavior}
\begin{figure}[h]
\centering	\includegraphics[width = 0.6\linewidth]{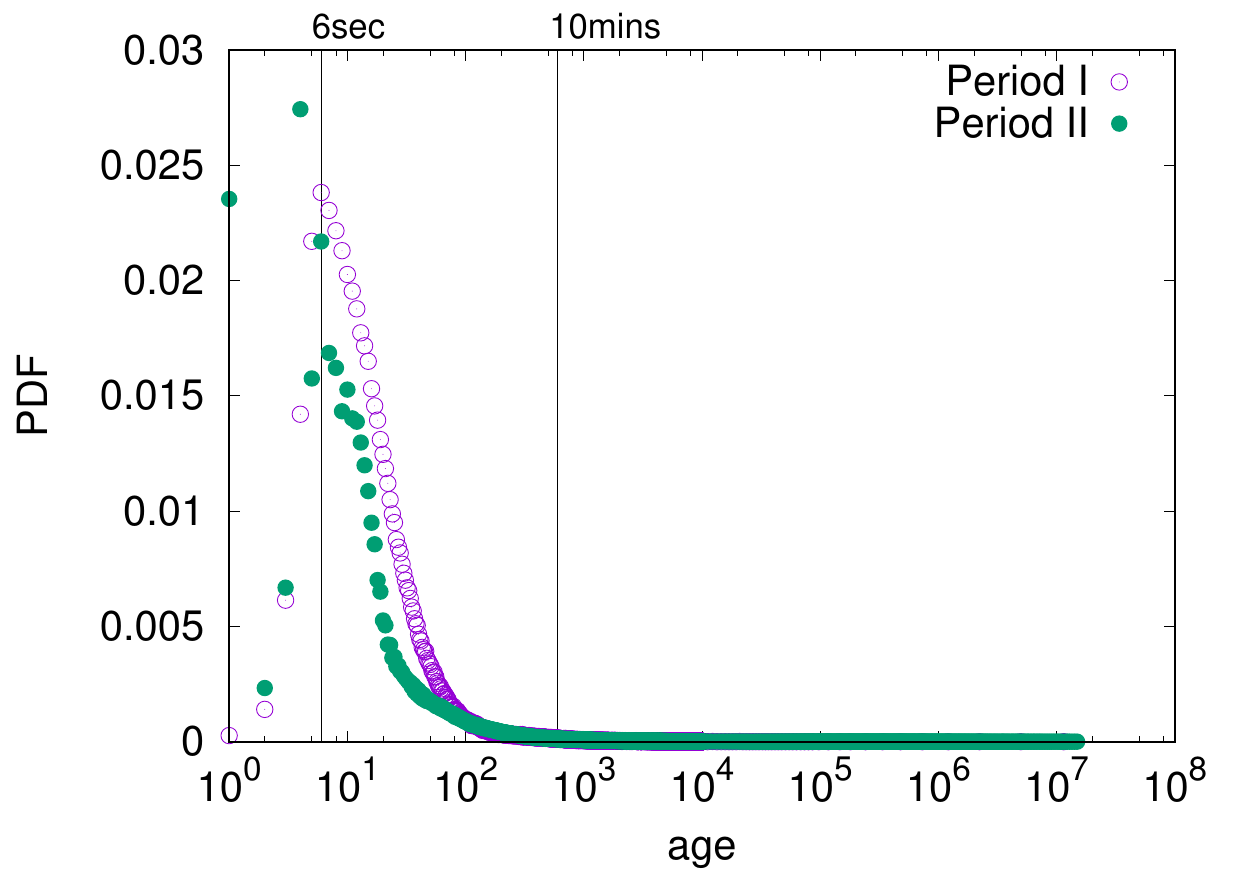}
	\caption{PDF of ages for posts with one comment for Periods I and II.}
	\label{fig:age_onecomment}
\end{figure}

Figure~\ref{fig:age_onecomment} shows the age distribution (frequency) of all the posts which have a single comment. In Period I, there is a very prominent peak at $6$ seconds, as found earlier (Figure~\ref{fig:age_posts}).
It can be seen that the ages of $72.78\%$ of these posts do not exceed $600$ seconds ($=10$ minutes). Period II looks very similar except the peak is seen at $5$ seconds with an additional peak at $1$ second.
\begin{table}[h]
	\begin{center}
		\caption{Cyborg-like Posts Statistics}
		\label{tab:Cyborg_data}
		\begin{tabular}{|l|r|r| }
        	\hline
        	  & Period I  & Period II \\
			\hline
			Posts with first comment in less than 6 seconds & 43138 & 1,804,374\\
			\hline
			Posts with same author of first comment & 7,615 & 492,928\\
			\hline
			Cyborg-like Posts & 6,389 & 387,845\\
			\hline
			Successful Cyborg-like Posts &  3,446 & 70,237\\
			\hline
			Successful Non Cyborg-like Posts & 866 & 28,892\\
			\hline
			Unsuccessful Cyborg-like Posts & 2,943 & 317,608\\
			\hline
            Unsuccessful Non Cyborg-like Posts & 360 & 76,191\\
            \hline
		\end{tabular}
	\end{center}
\end{table}
Furthermore, we analyzed posts whose first comment is posted within $6$ seconds, which constitutes $43138$ posts for Period I and $1,804,374$ for Period II. Out of these posts we found that  approximately $17\%$ and $20\%$ posts have their first comment by the author of the post for Period I and Period II respectively. To understand this behavior of posting comment by the same user, we checked the number of characters in the first comment of these posts. For instance, we find that $83.9\%$  and $79\%$ posts have number of characters more than $100$ for Period I and II respectively. It is crucial to mention that we have left out posts which contain links to web-pages in this analysis, which may be copied from a certain source and pasted in the posts.
Choice of such figures are based on the rationale that writing such long comments within $6$ seconds is quite impossible for a genuine human, and more likely to be done using automated means. Hence, we categorize these posts to be exhibiting a \textit{cyborg-like} behavior, where these posts may be just an advertisement or a message that these users intend to propagate.

\begin{figure}[h]
\centering	\includegraphics[width = 0.48\linewidth]{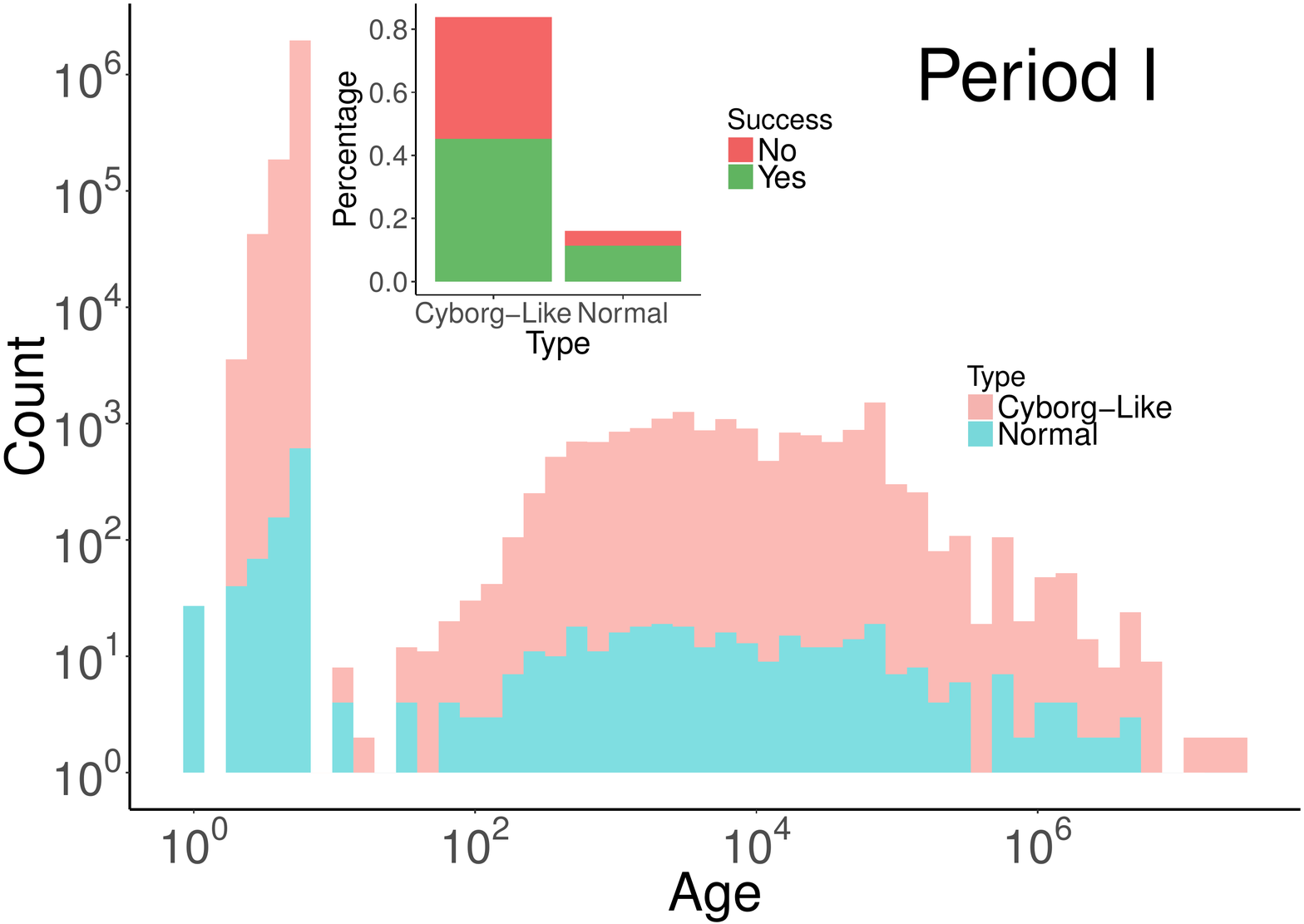}
\centering     \includegraphics[width = 0.48\linewidth]{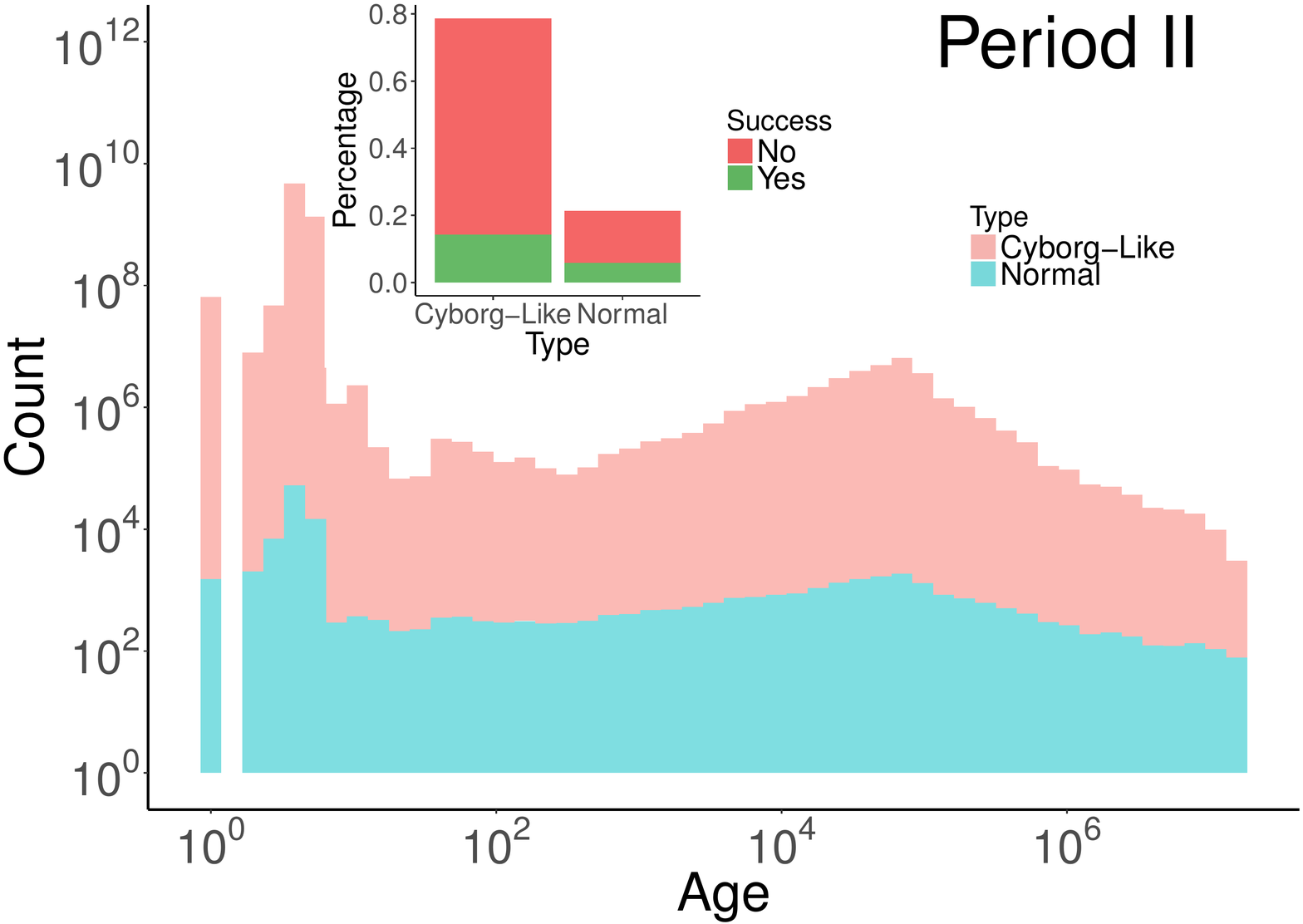}
	\caption{The histograms for the actual age of the posts for which the first comment arrived  within $6$ seconds. The insets show distribution of posts over success rate.}
	\label{fig:bot_inset}
\end{figure}
We also define a `success criterion' in order to check whether this type of posts were successful in garnering attention or responses from other users -- if a post is getting any reaction (comment or vote) from other Reddit users, then they are considered `successful' in drawing attention. We find that $53.93\%$ and $18\%$ \textit{cyborg-like} posts were successful in Periods I and II respectively. While $70.63\%$ of normal posts (which have comments with less than 100 characters) of Period I are successful, which can be assumed to be possibly done by humans (insets of Figure~\ref{fig:bot_inset}), which comes to about $27.5\%$ for Period II. Table~\ref{tab:Cyborg_data} summarizes the data for this analysis. 
Hence, for Period I, we infer that machine generated content is less likely to garner interest as compared to human generated content. It is reasonable to assume the low success of the cyborg-like posts are due to the fact that lengthy comments and promotions/advertisements provide less room for any discussions. 
For Period II the cyborg-like posts have smaller success rate, which requires a further granular analysis, due to the increased richness in the variety of behavior in the cyborg-like posts in recent times.

\subsection{Analysis of depth and breadth of a post}
\begin{figure}[h]
\centering \includegraphics[width = 0.48\linewidth]{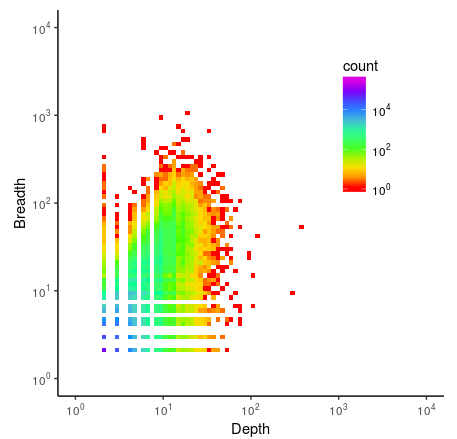}
\centering \includegraphics[width = 0.48\linewidth]{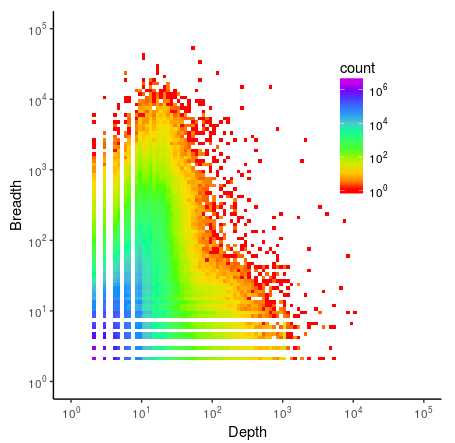}
		\caption{Plot showing the depth of the posts against the breadth of the posts, with the density of posts shown through the heat map.
	}
	\label{fig:DepthvsBreadth}
\end{figure}

Discussion happening on a post can be seen to have a tree-like structure. Depth of a post is defined by the maximum length of nested replies on a post and breadth of a post is defined as the maximum number of comments at a particular level. Figure~\ref{fig:DepthvsBreadth} shows the variation of depth against the breadth of the posts for both the periods, with the heat map depicting the density. 
We observe that depth and breadth can grow independent of each other. This is most prominent in Period II, where posts with simultaneously large values of depth and breadth are rare or absent.
The plot also shows that breadth grows more easily compared to depth, which can be attributed to the larger effort needed to grow or continue a nested discussion (increase depth) than to diversify a discussion by adding new parallel threads (increase breadth). 

{ The probability distributions of depth shows fat tails for both periods. Additionally, there are prominent peaks at certain specific values $676$ and $1000$ which account for several counting threads in counting subreddits where users incrementally count successive alphabet pairs (AA,AB, ...) and numbers upto $1000$ by replying to one another, and subsequently start a new thread when the series is complete. The probability distribution of breadth also shows fat tails with power law decay for the very high values. The extreme values correspond to subreddits named `Millionaire' where each user starts a new discussion at first level and thus contribute to increasing the breadth by unity.
}

{ 
\subsection{Dynamics of posts}
One can easily characterize a temporal sequence of events by computing the time difference $\tau$ between successive events.
Let us assume a sequence of events $n$, separated by a temporal distance $\tau_i, i=1,2,\ldots,n-1$.
One can define the burstiness of a signal as~\cite{goh2008burstiness}
\begin{equation}
    B = \frac{\sigma- \mu}{\sigma + \mu}
\end{equation}
where $\mu$ and $\sigma$ are the mean and standard deviation of a sequence of $\tau$.
By definition,  $B=1$ for the most bursty signal, $B=0$ for a neutral signal and $B=-1$ for a completely regular sequence.
We compute a sequence of $\tau$ and subsequently the value of the burstiness $B$.
This can be done for each author in terms of posts and comments, which corresponds to an author's posting burstiness or commenting burstiness. Additionally we can compute the burstiness of a post by considering all comments arriving in it.

\begin{figure}[h]
\includegraphics[width = 0.47\linewidth]{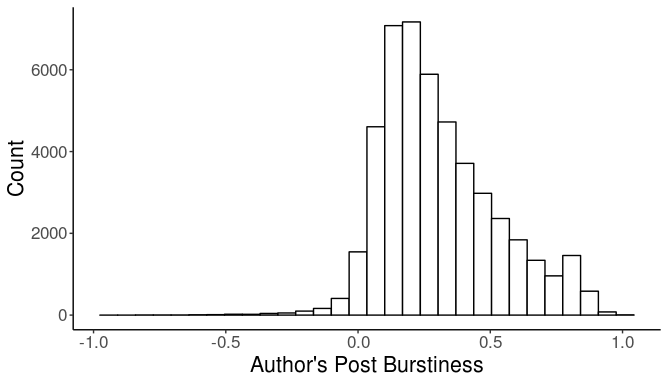}
\includegraphics[width = 0.47\linewidth]{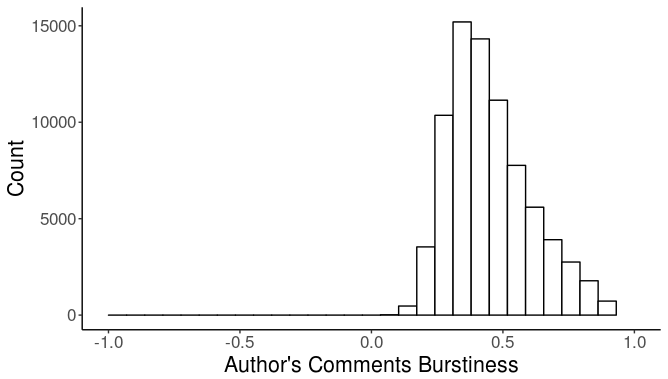}
\centering \includegraphics[width = 0.48\linewidth]{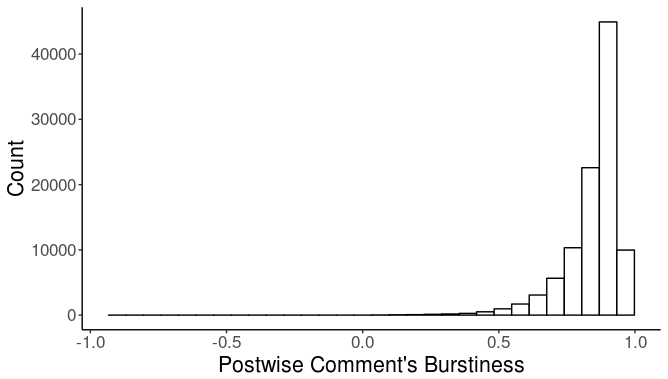}
		\caption{Plot showing the histogram of burstiness $B$ for 
		(i) authors' posting behavior (authors with at least 100 posts), 
(ii) authors' commenting behavior (authors with at least 500 comments),
and (iii) commenting behavior on posts (posts with at least 500 comments) 
for Period II. 
	}
	\label{fig:burstiness}
\end{figure}
Figure~\ref{fig:burstiness} shows the  histogram of burstiness $B$ for 
authors' posting behavior (authors with at least 100 posts), 
authors' commenting behavior (authors with at least 500 comments),
and commenting behavior on posts (posts with at least 500 comments) 
for Period II. 
We find that this distribution is skew with a larger fraction of posts with positive burstiness value ($B>0$), and in fact, the mean value $\tilde{B}$ is $0.31$ for authors' posting, $0.45$ for authors' commenting and $0.85$ for all comments on posts.
The first two quantities measure burstiness at the level of individuals and have seemingly similar distributions of burstiness and its mean, but the burstiness at the level of posts in terms of comments from various authors shows extremely skewed distribution of burstiness $B$ with very large average value as well.
This indicates that while activities at individual user level can be less bursty, the collective attention to a particular post can drive extremely bursty behavior.
This reaffirms the known hypothesis that human communication show bursty patterns~\cite{goh2008burstiness}.

 In real world communications, it is usual to find that the temporal spacing between successive human activities are quite heterogeneous, and range from completely random (following a Poisson distribution) to comments arriving at bursts.
}

\subsection{Popular Post Dynamics}
\begin{figure}[h]
\centering	\includegraphics[width = 0.48\linewidth]{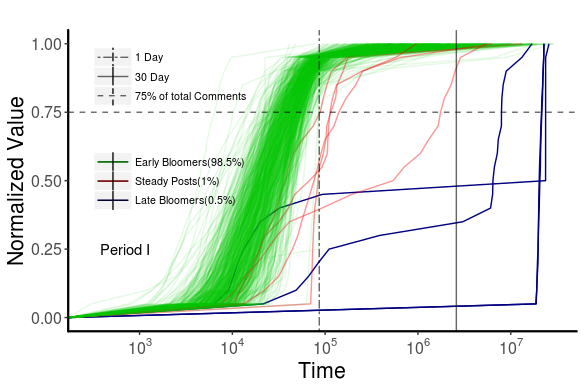}
\centering    \includegraphics[width = 0.48\linewidth]{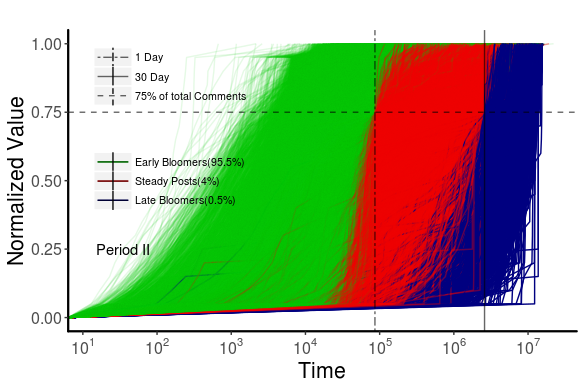}
		\caption{The time evolution of the number of comments in a post (normalized by the final number of comments obtained within our defined time window of 1 year) for all posts having more than $500$ comments. The data has been coarse-grained to aid visualization. The horizontal dotted line corresponds to $75\%$ of the final number of comments. The time to reach this fraction is used to characterize the posts. 
		The vertical lines are at 1 day and $30$ days. 
		Posts mostly active within 1 day  garner $75\%$ of their comments during that period (green).
		Some posts grow throughout their active life span taking time intermediate between 1 and $30$ days to reach the $75\%$ mark  (red), while others grow slowly while becoming active at some later stage, beyond the $30$ days period (blue). Plots are shown for Period I  and Period II.
        }
	\label{fig:comments_age_time1}
\end{figure}

To understand the age dynamics of the popular posts and infer their behavior, plot the time evolution of the posts which have more than $500$ comments. Three distinct categories are prominent (Figure~\ref{fig:comments_age_time1}):
\begin{itemize}
	\item \textit{Early bloomers} are rapidly growing posts, accumulating over $75\%$ of their total comments within $1$ day, creating the \textit{Mayfly Buzz} as discussed earlier,
	\item \textit{Steady posts} are characterized by steady activity throughout their lifespan.
	\item \textit{late bloomers} are slowly growing posts, which get suddenly very active at a late stage (after $30$ days). 
\end{itemize}
We also study the evolution of the total number of comments with the age of each post, for all posts in our data. The overlaid binned average of all data indicates a marked departure in the gross behavior around $1$ day.

\section{Analysis of interactions}
\label{sec:interaction}
The Reddit post-comment structure constitutes a tree graph, where the starting node is a post, and it can have its comments, and the comments can further garner replies. We defined a \textit{limelight score} for each post based on the number of comments gathered as reply to a single first-level comment. In a way, this score can compute the depth of discussion around a single comment for a given post. 
\begin{equation*}
\textrm{Limelight Score} = \frac{\max(Comm_j)}{\sum_{k = 1}^{N} Comm_j},
\end{equation*}
$Comm_j$ being the total number of comments under $j^{th}$ first level comment and $N$ is the total number of first level comments for that post.

\begin{figure}
 \centering   \includegraphics[width = 0.48\linewidth]{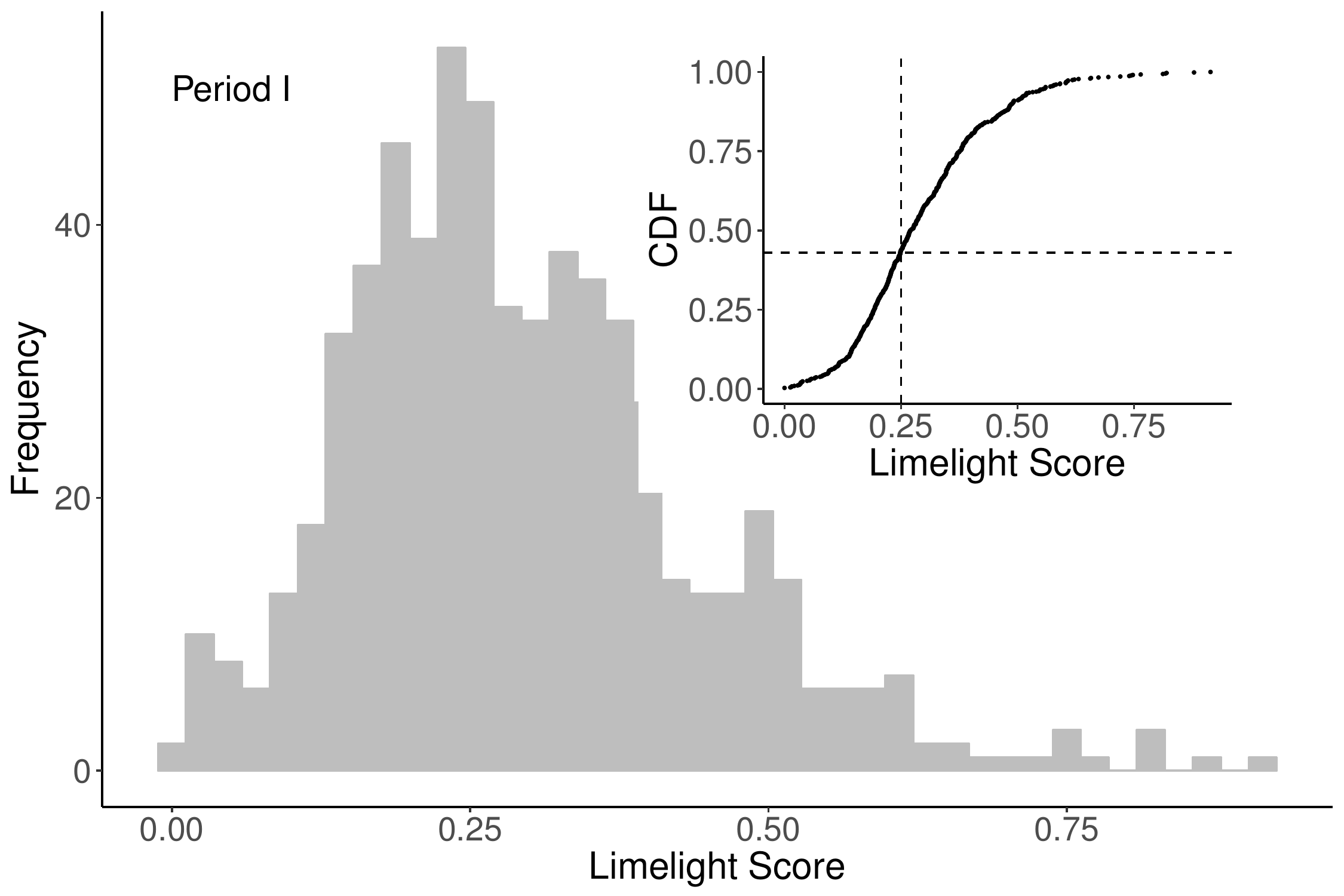}
 \centering   \includegraphics[width = 0.48\linewidth]{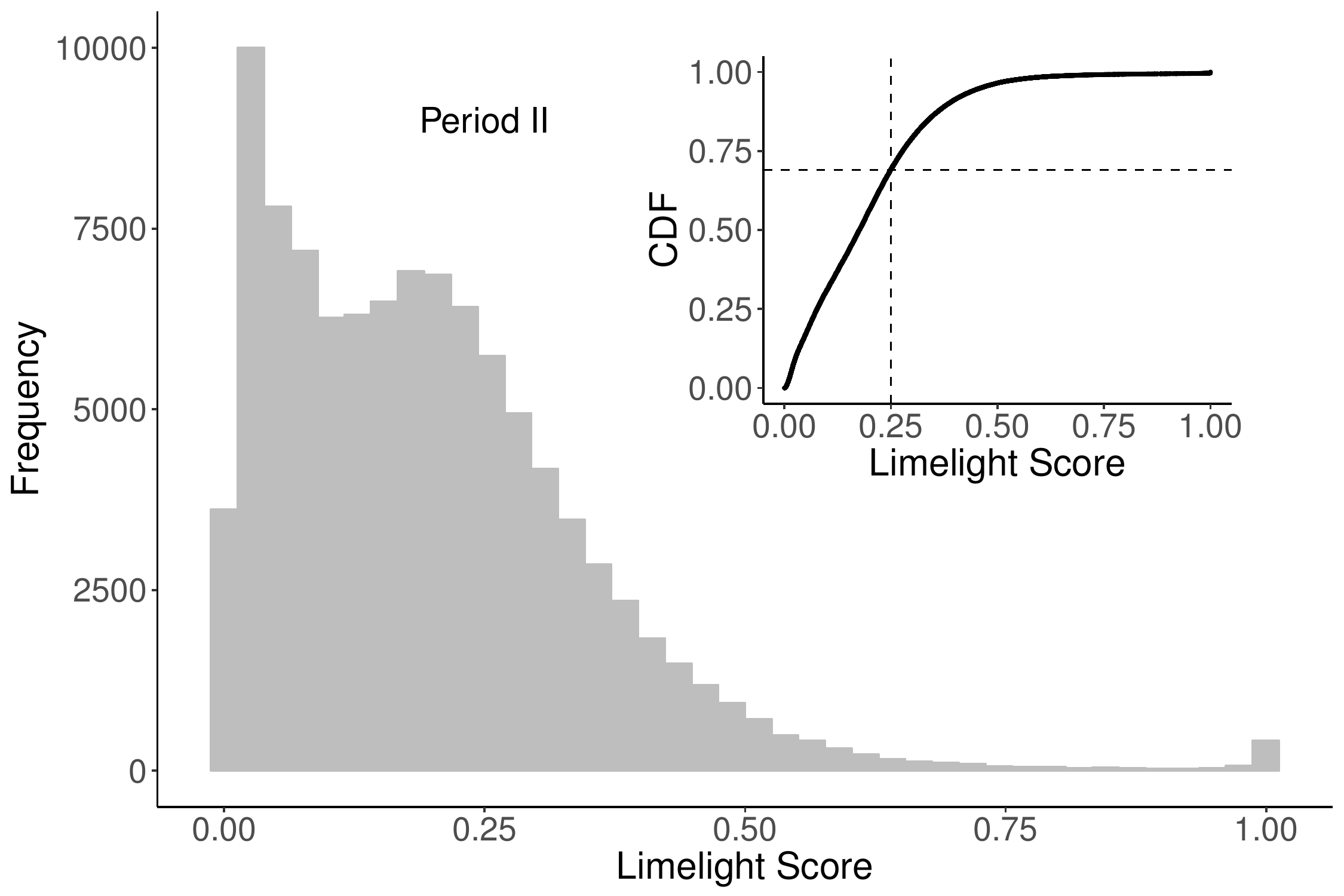}
    \caption{Histograms for Limelight scores for all posts having at least $500$ comments in Period I and Period II. The corresponding CDFs are shown in insets.}
    \label{fig:limelight}
\end{figure}
Figure~\ref{fig:limelight} shows the histogram of the \textit{Limelight scores} and the inset shows its CDF. We have considered posts that have at least 500 comments. 
We find that in Period I, $56\%$ of the total posts contain one comment with \textit{Limelight score} of at least $0.25$, i.e., at least 25\% of the discussion in this post is initiated and centered around a single comment.
Similar behavior is also present in $31\%$ of the posts in Period II.  Additionally in Period II, a finite number of posts actually have Limelight score close to unity, indicating absolute dominance of one branch of the comment tree for those posts.
We further observe that most of the time, the author of the post and that of the first level \textit{Limelight} hogging comment are not the same. This holds for about $97\%$ of the posts during Period I, for instance. 

This leads to a very interesting insight that links virtual human behavior in social media to physical world social behavior. It is quite common scenario that in course of any group discussion, usually  a few specific people apart from the presenter who pro-actively initiates a conversation by asking a question or making a comment, following which other people join the conversation by making comments or replies. Surprisingly, it is observed that lime-light hogging behavior is completely missing for posts whose authors exhibit \textit{Cyborg-like} behavior. 
Thus, it can be inferred that posts automatically generated by bots have failed to garner human attention most of the times. However, rigorous studies can be undertaken in future to validate this inference.

To the best of our knowledge, characterizing content popularity by the depth of discussion around it has not been attempted before. Since it has been proved in earlier studies~\cite{stoddard2015popularity,glenski2017consumers} that the number of upvotes-downvotes are not meaningful indicators for measuring interestingness or popularity of content, we claim that this can be a good way to measure them.

\section{Analysis of Author Behavior}
\label{sec:author}
To analyze author interactions, we started by defining a network with nodes representing unique authors and edges representing the interaction between the authors using comments. We define the in-degree and out-degree for each node based on the number of interactions, where a self loop is ignored. The gross statistics for the $3$ categories of authors -- (i) who only put up posts but don't comment on others' posts are the pure \textit{content producers}, (ii) who only comment are the pure \textit{content consumers}, and (iii) the rest of them indulge in both posting and commenting, are summarized in Table~\ref{tab:author_data}.

\begin{table}[h!]
	\begin{center}
		\caption{Author Table}
		\label{tab:author_data}
		\begin{tabular}{|l|r|r|}
        	\hline
        	  & Period I & Period II \\
			\hline
			Total Active Authors & 229,488 & 9,369,708\\
			\hline
			Total Authors who only create posts & 140,918 & 1,917,161\\
			\hline
			Total Authors who only comments & 39,764 & 3,019,676\\
			\hline
			Total Authors who comment as well create posts & 48,806 & 4,432,871\\
			\hline
		\end{tabular}
	\end{center}
\end{table}

We try to quantify author interactions to assess their influence.
If the total effective number of comments received is given by $A$  and the total number of comments on others' posts is given by $B$, then we can define the \textbf{interaction score} of an author as $A/(A+B)$. This score is trivially zero for all authors who comment on others' posts but have not received any comments on their posts. Score is trivially $1$, for an author who does not comment on others' posts but receives comments on one's own posts. This is rather rarely observed. For both periods, peaks at 0, 0.5 and 1 are prominent.



There are some distinct authors who have the ability to consistently garner a large number of comments on each of their posts. We analyze the average number of effective comments received per post by authors, in order to quantify this. 
We observed that $22\%$ of the authors in Period I and $6\%$ in Period II have fewer effective comments than the number of posts that they have put up which corresponds no interaction for many posts. $11\%$ in Period I and $13\%$ in Period II have received one comment per post on the average. The rest, amounts to $67\%$ for Period I and $81\%$ for Period II received more comments than the number of posts put up for the respective periods.

The discussion above indicates that authors who receive more attention on their posts are seemingly the ones who comment on others' posts. Simply put, in order to gain attention on social media, authors need to be reciprocative in nature, which is also indicated by the peak at interaction score  of $0.5$.
It also emphasizes the fact that the social media interactions are dominated by mutual gratification.

\section{On signatures of controversies}
\label{sec:controversy}
{ Contrary to the graph based approaches common in the literature , which deal with controversies in social media platforms (see, e.g., Ref.~\cite{garimella2018quantifying}), we take a rather moderate, statistical approach, which attempts to lay down the foundation for a rigorous, sentiment analysis based approach in the future.}

\begin{figure}[t]
\centering	\includegraphics[width = 0.67\linewidth]{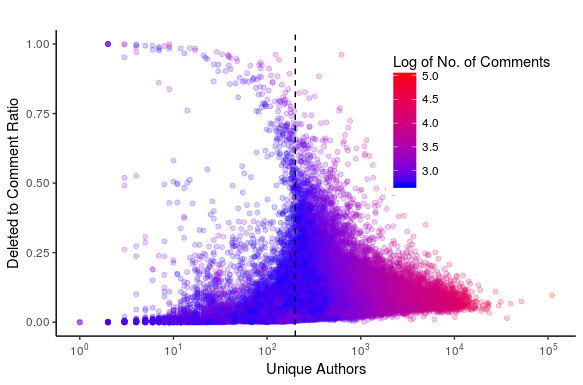}
\centering	\includegraphics[width = 0.67\linewidth]{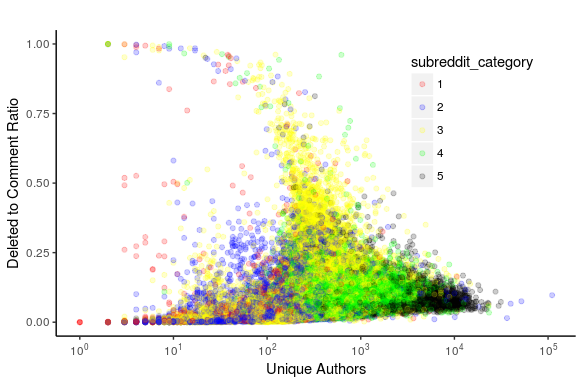}
	\caption{(Top) The plot of controversiality score against the number of unique authors in the post, for all posts in Period II with at least 500 comments. Each post is coloured according to the number of comments it has.
	(Bottom) Same plot where the color of the point is according to the category of the subreddit in which they belong (discussion in text).}
	\label{fig:uniqueauthdelratio}
\end{figure}

We observe that in the popular posts (with more than $500$ comments), some 
comments have been deleted either by the author of the comment or by the moderators of the subreddits. In the latter case, deletion of a comment can happen only if the author made a comment that violates the rules of the subreddit set by the moderators, that can potentially lead to controversy in a social discussion platform. For further analysis, we have calculated the ratio of the number of deleted comments to the total number of comments, which can serve as a proxy for the measure of controversiality of a post and call it the \textit{Controversiality Score}. In the top panel of  Figure~\ref{fig:uniqueauthdelratio} we plot the Controversiality Score of a post against the number of unique authors for Period II. We observe that the plot branches roughly into two components for lower number of unique authors, one each for very high and very low level of controversiality score. This branching is absent beyond a certain number of unique authors, which is roughly $200$ for our case. In the top panel of Figure~\ref{fig:uniqueauthdelratio}, the colors map to the number of comments in the posts. 

We observe that when there are fewer unique authors (less than $200$ in our case) contributing in a discussion to a post, the outcomes can be quite extreme -- it can either see a very high degree of controversy or a very low degree of controversy, as is seen by the left part of the graph. Beyond $200$ unique authors, this extreme diversity vanishes -- in fact, very high values of controversiality are absent. This indicates that more controversiality occurs within smaller groups than over larger ones.

\begin{figure}[h]
\centering	\includegraphics[width = 0.85\linewidth]{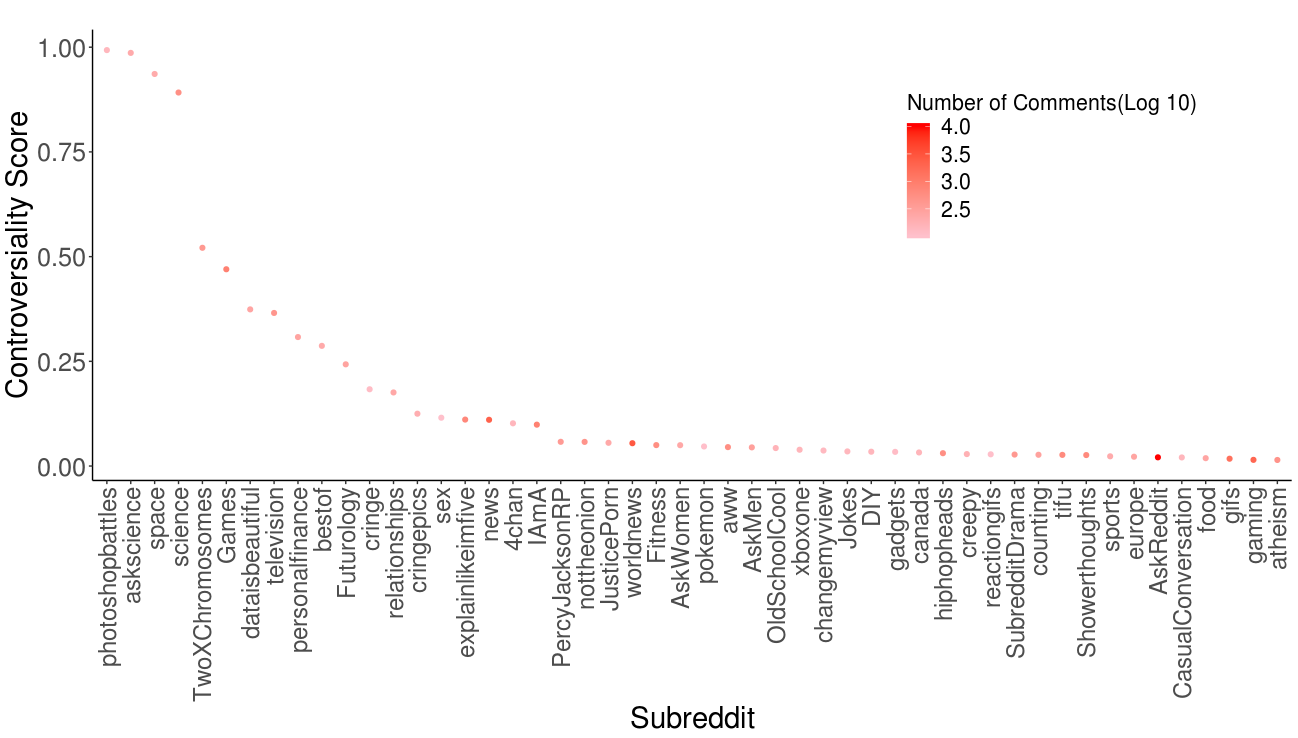}
\centering	\includegraphics[width = 0.85\linewidth]{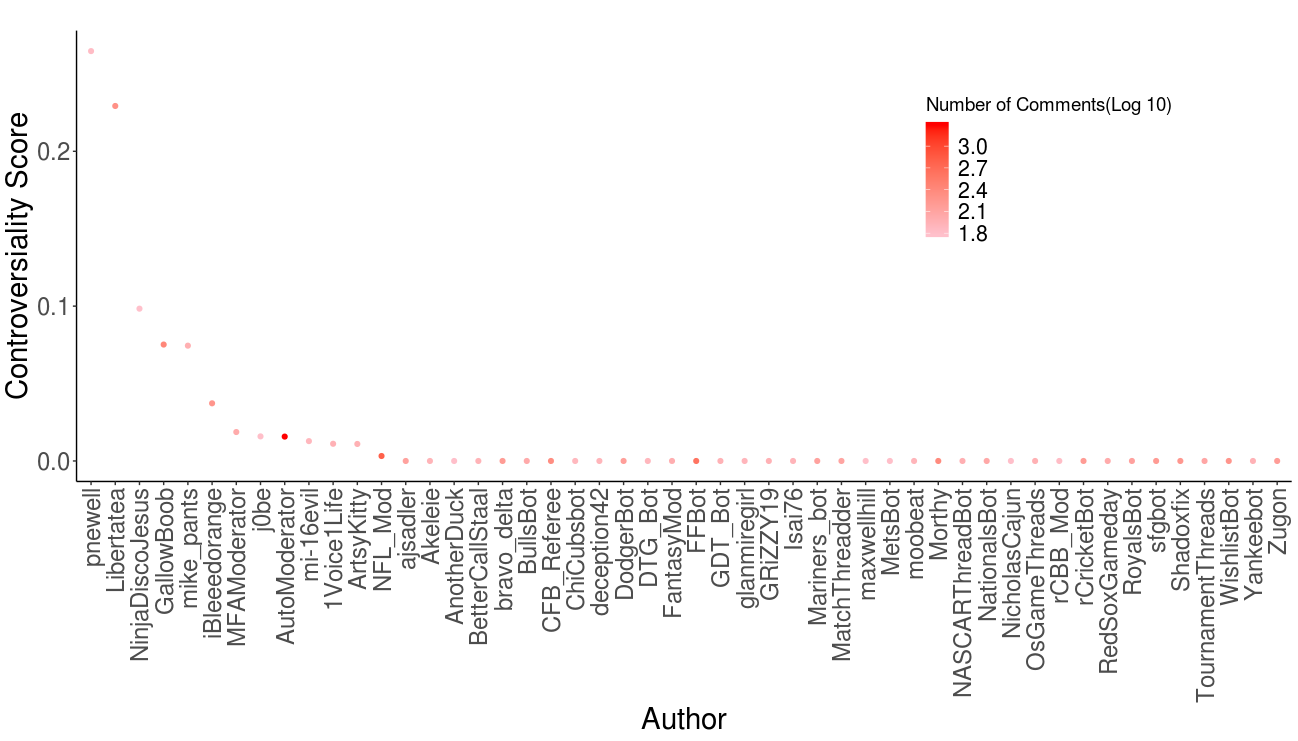}
	\caption{(Top) Plot for the controversiality score for the subreddits which have more than $100$ posts.
	(Bottom) Plot for the controversiality score for the authors which have more than $50$ posts.}
	\label{fig:subredditcontroversiality}
\end{figure}
We further wanted to check if controversiality is related to the popularity of the subreddit in which the post is created.
We define the popularity of a subreddit as the total number of posts that are being created in that subreddit during the period, and divide them into $5$ categories, $1$ being the least popular subreddit and $5$ being the most popular subreddit (1 - $1-10$ posts, 2 - $11-100$, 3 - $101-1000$, 4 - $1001-2000$, 5 - above $2000$). 
In the bottom panel of Figure~\ref{fig:uniqueauthdelratio}, we plot the controversiality score of posts against the number of unique authors with colors indicating the popularity category of the subreddit. We can see that the popular subreddits (categories 4 and 5) have low controversiality score, around $0.25$  or less, while high controversiality scores are prevalent in the less popular categories.
Hence, we infer that controversiality is observed in smaller, closely knit groups (akin to \textit{contempt breeds contempt}), and users stay clear of controversies in larger groups.

We have also checked the controversiality of the individual subreddits. We consider subreddits which have at least $100$ posts. 
A post is considered \textit{controversial} if its controversiality score is more than $0.2$ i.e., more than $20\%$ of comments of that posts are deleted. We calculated controversiality score of a subreddit as the fraction of posts in it that are controversial. The top panel of  Figure~\ref{fig:subredditcontroversiality} shows the controversial score of those subreddits in which users have posted at least $100$ posts.

To check which users are responsible for the above, we can compute author-wise controversiality score. We can extend the above definition to  \textit{author controversiality score} which is the ratio of the number of controversial posts to the total number posts by the author where controversial posts are those with more than $20\%$ percent of deleted comments in the posts. The bottom panel of Figure~\ref{fig:subredditcontroversiality} shows the author controversiality score of authors who have more than $50$ posts in our data.

The above measures are the indicators that tell us in which Reddit community (subreddit) controversial posts are put up and which user is responsible for initiating the controversy.

\section{Conclusions and outlook}
\label{sec:conclusion}

A large, community-driven social network and discussion platform like Reddit harbors a plethora of behaviors for the users concerned. A huge fraction of posts are left uncommented, while some gather of a considerable amount of attention through actions on them like comments and votes. The distribution of the number of comments on posts show correlation through the power law tail~\cite{thukral2018analyzing}, and the behavior of authors show a large variation -- while many authors simultaneously create post and write comments, there are also a large fraction of purely \textit{content producers} and \textit{content consumers}, who restrict themselves only to either  posting and commenting respectively.
{  The distribution of the number of comments by unique authors exhibit
lognormal distribution for the largest values, which indicates an underlying multiplicative process, and thus a strong correlation between authors.  } 
Each post stays active as comments are added and thus discussions are produced. However, a huge fraction of posts are left with only a single comment, and within them, a majority receive that only single comment within $6$ seconds, containing a large number characters unlikely to be written by an average human, indicating a \textit{Cyborg-like} behavior. Further to that, a large fraction of posts become inactive around the age of $1$ day. This is consistent with the average active time of posts reported for micro-blogging site such as Twitter~\cite{kwak2010twitter}. When we studied the time evolution of the top commented posts, we found three broad classes for the posts -- (i) \textit{early bloomers} who gather more than $75\%$ of their lifetime comments within a day, (ii) \textit{steady posts} whose number of comments grow steadily throughout their lifespan, and (iii) \textit{late bloomers} who show very little activity until steadily gathering comments near the end of their lifespan. The early bloomers contribute to what we term as \textit{Mayfly Buzz}, and constitute the majority of the posts.
Posts also show \textit{limelight hogging} behavior and we find that $56\%$ of posts in Period I and $31\%$ of posts in Period II have \textit{limelight score} above $0.25$, indicating that for a large fraction of posts, at least one-fourth of the total weight of the discussions are contributed by one chain of comments.
In fact, this measure can be a more meaningful indicator of the  interestingness or popularity of the content, compared to just votes or only number of comments.
Social media discussion threads sometimes contain controversial content, and in Reddit this is  moderated by deleting posts or comments. Our study tries to measure the controversiality from the fraction of such deletions, at the level of posts, authors and subreddits. We observe that controversiality is more prevalent in small, closely knit groups than large ones.
Analysis of actual content can lead us to a better understanding, which we plan to carry out in future studies.
{ Our initial measurements of sentiment on the text content of comments around deletions did not indicate any significant signal, but probably a further careful analysis of other quantities along with sentiment can help us formulate a unique indicator which can eventually be used for prediction/forecasting of unruly textual events.

We also investigated the temporal patterns of events, in terms of the posts created by individual authors, the comments created by individual authors, as well as the comments on popular posts. All of them show bursty behavior of events, validating the fact that human communication is usually bursty in nature.
}

With the increasing use of social media even within closed groups as well as organizations, understanding human behaviors and ability to characterize them is turning out to be an important task with potential impact and applications. One possible application of understanding temporal patterns of group behavior in such a scenario can be focused on injecting the right content or advertisement for the right group at the right time.

Our analysis brings out a variety of behavioral elements from the authors and through their interactions. There are few authors who are able to generate quite a lot of activity across a large number of posts. Going ahead, analysis of change of sentiment can turn out to be interesting. The insights gained from this analysis can be used to model different aspects from a large interactive population. In addition, predicting the recent trends can lead to better targeted reach e.g., innovative usage of \textit{memes} etc.

\bibliographystyle{splncs04}
\bibliography{sample-bibliography}	

\end{document}